\documentclass[conference]{IEEEtran}
\IEEEoverridecommandlockouts

\usepackage[utf8]{inputenc} 
\usepackage[T1]{fontenc}
\usepackage{url}
\usepackage{cite}
\usepackage[cmex10]{amsmath} %
\usepackage{amssymb, amsfonts}
\usepackage{aligned-overset}
\usepackage{graphicx}         %
\usepackage[nolist]{acronym} 
\usepackage[normalem]{ulem}
\usepackage{nicematrix}
\usepackage{booktabs}
\usepackage{tabularx}
\usepackage{multirow}

\usepackage{tikz}
\usetikzlibrary{arrows,calc,shapes,graphs,graphs.standard,quotes,arrows.meta,decorations.markings,positioning,fit,positioning,hobby,decorations.text,decorations.pathreplacing}

\usepackage{pgfplots}
\pgfplotsset{compat=1.18}
\pgfdeclarelayer{background}
\pgfdeclarelayer{foreground}
\pgfsetlayers{background,main,foreground}
\usepgfplotslibrary{groupplots}

\usepackage[hidelinks]{hyperref}
\usepackage[capitalise]{cleveref}
\usepackage{amsthm}

\crefname{proposition}{Proposition}{Propositions}
\newtheorem{proposition}{Proposition}%
\newtheorem{definition}{Definition}%

\begin{acronym}
\acro{AWGN}{additive white Gaussian noise}
\acro{BDD}{bounded distance decoding}
\acro{BCH}{Bose-Ray-Chaudhuri-Hocquenghem}
\acro{BEC}{Binary Erasure Channel}
\acro{BER}{bit error rate}
\acro{BMI}{bitwise mutual information}
\acro{BPSK}{binary phase shift keying}
\acro{CP}{Chase-Pyndiah}
\acro{GRAND}{guessing random additive noise decoding}
\acro{HD}{hard decision}
\acro{HDD}{hard decision decoding}
\acro{iBDD}{iterative bounded-distance decoding}
\acro{LLR}{log-likelihood ratio}
\acro{MAP}{maximum a posteriori}
\acro{MI}{mutual information}
\acro{ML}{maximum likelihood}
\acro{ORBGRAND}{ordered reliability bits GRAND}
\acro{PC}{product code}
\acro{SD}{soft decision}
\acro{SDD}{soft decision decoding}
\acro{SISO}{soft-in/soft-out}
\acro{SOCCP}{soft-output from constant codebook probability}
\acro{SOCS}{soft-output from covered space}
\acro{SOGRAND}{soft-output guessing random additive noise decoding}
\acro{TPD}{turbo product decoding}
\acro{SNR}{signal-to-noise-ratio}
\end{acronym}
\definecolor{mittelblau}{RGB}{0, 126, 198}
\definecolor{violettblau}{cmyk}{0.9, 0.6, 0, 0}
\definecolor{rot}{RGB}{238, 28 35}
\definecolor{apfelgruen}{RGB}{140, 198, 62}
\definecolor{gelb}{RGB}{255, 229, 0}
\definecolor{orange}{RGB}{244, 111, 33}
\definecolor{pink}{RGB}{237, 0, 140}
\definecolor{lila}{RGB}{128, 10, 145}
\definecolor{hellgrau}{RGB}{224, 224, 224}
\definecolor{mittelgrau}{RGB}{128, 128, 128}
\definecolor{dunkelgrau}{RGB}{80,80,80}
\definecolor{anthrazit}{RGB}{19, 31, 31}
\definecolor{aqua}{RGB}{0, 255, 255}
\renewcommand{\mid}{\,|\,}

\newcommand{\vL}{\boldsymbol{L}}

\newcommand{\vc}{\boldsymbol{c}}

\newcommand{\vn}{\boldsymbol{n}}

\newcommand{\vu}{\boldsymbol{u}}
\newcommand{\vv}{\boldsymbol{v}}

\newcommand{\vy}{\boldsymbol{y}}

\DeclareMathOperator{\MI}{I}

\newcommand{\Cod}{\mathcal{C}}
\newcommand{\Codi}{\mathcal{C}_i}
\newcommand{\dmin}{d_{\mathrm{min}}}
\newcommand{\List}{\mathcal{L}}
\newcommand{\Listi}{\mathcal{L}_i}
\newcommand{\Ind}{{\mathcal{I}}}
\newcommand{\Kcal}{\mathcal{K}}
\newcommand{\one}{\boldsymbol{1}}
\newcommand{\pws}{\mathcal{P}}

\newcommand{\Px}{P}
\newcommand{\Pyx}{p}
\newcommand{\Pcy}{\Px}
\newcommand{\Pvcy}{\Px}
\newcommand{\Pvvy}{\Px}

\newcommand{\Lch}{L^{\mathrm{ch}}}
\newcommand{\Lin}{L^{\mathrm{in}}}
\newcommand{\Lapp}{L^{\mathrm{app}}}
\newcommand{\Lext}{L^{\mathrm{ext}}}
\newcommand{\chd}{\hat{y}}

\newcommand{\vLin}{\vL^{\mathrm{in}}}
\newcommand{\vchd}{\hat{\vy}}

\newcommand{\Ball}{\mathcal{B}}
\newcommand{\Ftwo}{\mathbb{F}_{2}}
\newcommand{\Vcov}{\mathcal{V}}
\newcommand{\Vld}{\Vcov_{\mathrm{LD}}}
\newcommand{\ttest}{\boldsymbol{\tau}}
\newcommand{\Test}{\mathcal{T}}
\newcommand{\Nat}{\mathbb{N}}

\newcommand{\blackdot}{\tikz[baseline=-0.6ex] \fill (0,0) circle (2.2pt);}
\newcommand{\greydot}{\tikz[baseline=-0.6ex] \draw[black] (0,0) circle (2.2pt);}

\tikzset{cross/.style={cross out, draw, 
     minimum size=4pt, 
     very thick,
     inner sep=0pt, outer sep=0pt}}
\newcommand{\redcross}{\tikz[baseline=-0.5ex] \node[cross, draw=rot] {};}
\newcommand{\bluecross}{\tikz[baseline=-0.5ex] \node[cross, draw=mittelblau, rotate=45] {};}

\newcolumntype{C}[1]{>{\centering\arraybackslash}p{#1}}
\newcommand{\STAB}[1]{\begin{tabular}{@{}c@{}}#1\end{tabular}}

\begin{document}

\title{Soft-Output from Covered Space Decoding\\ of Product Codes
}

\author{\IEEEauthorblockN{Tim Janz$^{\ast}$, Simon Obermüller$^{\ast}$, Andreas Zunker, and Stephan ten Brink}
\IEEEauthorblockA{Institute of Telecommunications, 
University of Stuttgart, Germany\\
\{janz, obermueller, zunker, tenbrink\}@inue.uni-stuttgart.de}
\thanks{$^{\ast}$These authors contributed equally to this work.}
}

\maketitle

\begin{abstract}
In this work, we propose a new soft-input soft-output decoder called \ac{SOCS} decoder. 
It estimates the a posteriori reliability based on the space explored by a list decoder, i.e., the set of vectors for which the list decoder knows whether they are codewords.
This approach enables a more accurate calculation of the a posteriori reliability and results in gains of up to 0.25\,dB 
for turbo product decoding with \ac{SOCS} compared to \acl{CP} decoding.
\end{abstract}


\section{Introduction}\label{sec:intro}

\Acp{PC}\cite{elias1954} and their spatially coupled generalizations \cite{smith2012}\hspace{1sp}\cite{shehadeh2025} are a popular family of codes for high-speed optical communications. 
They achieve high net coding gains with low-complexity constituent decoders. 
The component codes allow for \acf{iBDD} based on efficient algebraic decoders, often referred to as \acf{HDD}. To achieve higher coding gains, \acf{SDD} can be used, which exploits the soft information of the channel. Typically, \ac{TPD} using \ac{SDD} for \acp{PC} is performed with the \acf{CP} decoder \cite{pyndiah1998}.

Several schemes have been proposed that use soft-in\-for\-ma\-tion while significantly reducing the data flow between the decoders compared to \ac{CP} decoding.
One approach is to use the most reliable bits as anchors to avoid miscorrections \cite{hager2018}.
Other promising approaches coarsely quantize soft-information \cite{lei2019, sheikh2019a, miao2022}. All of them improve upon \ac{iBDD} closing the gap to \ac{CP} decoding.
In \cite{strasshofer2023} and \cite{deng2023}, the soft-output of the \ac{CP} decoder is improved by optimizing parameters or adapting them dynamically, respectively.

Recently, in \cite{yuan2025a}, the authors proposed a novel approach to generate more accurate a posteriori \acp{LLR} compared to \ac{CP} decoding.
First, a list of candidate codewords is found using \ac{GRAND}. 
Then, all the noise queries made by \ac{GRAND} to obtain the list are used to estimate the probability that the correct codeword is not in the list.

Inspired by this idea, we propose a novel soft-input soft-output decoder that estimates the codebook probability using the code structure and the space covered by the constituent list decoder. 
It approximates the probability of all codewords that are not in the list using vectors in Hamming balls around the testwords or the codewords of the list.
Our proposed decoding algorithm, called \emph{\acf{SOCS}} decoding, performs up to $0.25\,\mathrm{dB}$ better than \ac{CP} decoding for high-rate product codes.
Thus, the proposed algorithm reduces the gap between \ac{CP} decoding and typically infeasible iterative decoding with optimal a posteriori component decoders.

\section{Preliminaries}\label{sec:prelim}
For $n\in \Nat$, let $[n]\triangleq\{1,2,\ldots,n\}$. 
The power set of an index set $\Kcal \subseteq [n]$ is denoted by $\pws_{\Kcal}$ with $|\pws_{\Kcal}|=2^{|\Kcal|}$.
Let $\vv_\Kcal\in\Ftwo^{|\Kcal|}$ be the subvector of $\vv \in \Ftwo^n$ containing only the elements at positions $\Kcal \subseteq [n]$ of $\vv$.
The vector of length $n$ with ones at positions $\Kcal\subseteq [n]$ and zeros at positions $[n]\setminus \Kcal$ is denoted by $\one_{\Kcal}$.
For two vectors $\vu,\vv\in\Ftwo^n$, let $\vu \oplus \vv$ be the element-wise addition over $\Ftwo$. 
The Hamming ball of radius $r$ centered at $\vu$ is denoted by $\Ball_r(\vu) \triangleq \{\vv \in \Ftwo^n \mid d_\mathrm{H}(\vu, \vv) \leq r\}$.
Similarly, the $(n-\left\lvert\Kcal\right\rvert)$-dimensional Hamming ball in $\Ftwo^n$ with respect to positions $[n] \setminus \Kcal$ is $\Ball_r(\vu,\Kcal) \triangleq \{\vv \in \Ball_r(\vu)\mid \vv_\Kcal = \vu_\Kcal\}$.
For a set of vectors $\Vcov$, we define the union of Hamming balls as $\Ball_r(\Vcov) \triangleq \bigcup_{\vv \in \Vcov} \Ball_r(\vv) $ and $\Pcy(\Vcov\mid \vy) \triangleq \sum_{\vv\in\Vcov} \Pcy(\vv\mid \vy)$.

\subsection{Channel Model}\label{subsec:channel}

We assume transmission over a binary-input additive white Gaussian noise channel $Y=X+Z$, where $Z\sim \mathcal{N}(0,\sigma^2)$ and $X\in\{+1,-1\}$ is the \ac{BPSK} modulated channel input with $0\mapsto+1$ and $1\mapsto-1$. 
We can compute the channel \ac{LLR} for a given channel output $y$ as
\begin{equation*}
    \Lch = \ln\frac{\Pyx(y\mid x=+1)}{\Pyx(y\mid x=-1)} = \frac{2}{\sigma^2} \, y.
\end{equation*}
The \emph{a posteriori} probability that the hard decision 
$\chd\in \Ftwo$, where $\chd = 1$ if $y < 0$ and $0$ otherwise,
equals the transmitted bit $v \in \Ftwo$ is given by
\begin{equation*}
    \gamma\triangleq \Pcy(v=\chd\mid y) = \frac{1}{1+\mathrm{exp}(-|\Lch|)},
\end{equation*}
while the probability that the hard decision is incorrect is given as $1-\gamma = \Pcy(v\neq\chd\mid y)$.
The probability that a vector ${\vv\in \Ftwo^n}$ was transmitted,
assuming that the elements of $\vv$ are uniform and i.i.d.,
can then be computed by
\begin{equation}
    \Pvcy(\vv\mid\vy) = 
    \prod\limits_{i:\,v_i=\chd_i}\gamma_i
    \cdot\prod\limits_{i:\,v_i\neq\chd_i} (1-\gamma_i).
\label{eq:cwpros_from_gamma}
\end{equation}

\subsection{Product Codes}\label{subsec:PC}
Let $\Cod\subseteq\Ftwo^n$ be an $(n,k,\dmin)$ binary linear code where $n$, $k$, and $\dmin$ denote the length, dimension, and minimum Hamming distance of the code, respectively. 
We consider two-dimensional \acp{PC} where the constituent codes for rows and columns are identical. 
A codeword of the \ac{PC} with parameters $(n^2,k^2,\dmin^2)$ and rate $R=k^2/n^2$ can be represented by an $n\times n$ array, where each row and column is a codeword of $\Cod$.

\subsection{Chase-II Algorithm}

A widely used soft-input list decoder is the Chase-II algorithm \cite{chase1972}. 
The algorithm processes the incoming message vector $\vLin$ as follows. 
First, the $p$ indices
\begin{equation*}
    \Ind = \mathop{\arg \min}\limits_{\Kcal \subseteq [n],\,\lvert\Kcal\rvert=p}\; \sum\limits_{i \in \Kcal} \left\lvert\Lin_i\right\rvert
\end{equation*}
corresponding to the least reliable positions of $\vLin$ are identified and a hard decision vector $\vchd = (\chd_1, \ldots, \chd_n)$ is obtained.
From $\vchd$, a set of $2^p$ testwords $\Test = \left\{\vchd \oplus \one_{\Kcal}\;\middle|\; \Kcal \in \pws_{\Ind}\right\}$ is generated.
Each testword is then decoded using \ac{BDD}.
All unique codewords from the decoding attempts form a list $\List$ of candidate codewords.

\subsection{Turbo Product Decoding}\label{subsec:pyndiah_strasshofer}

\Acp{PC} can be decoded with \ac{TPD} that iteratively improves the a posteriori reliability of each position using alternately row and column constituent decoders.
Given the channel output $\vy$, the optimal constituent decoder%
calculates the a posteriori \ac{LLR} by
\begin{equation}
    \Lapp_i = \Lext_i + \Lin_i = \ln\frac{ \Pvcy(\Codi^0\mid\vy)}{ \Pvcy(\Codi^1\mid\vy)},
\label{eq:lapp_def}
\end{equation}
where $\Codi^s = \left\{\vc \in \Cod\,\middle|\, c_i=s\right\}$ denotes the sets of codewords that are $s\in\Ftwo$ at the $i$th position, respectively.

Suboptimal constituent decoders approximate the a posteriori \ac{LLR}, typically using heuristic parameters and post-processing steps, e.g., as in \ac{CP} decoding \cite{pyndiah1998}. To optimize these parameters efficiently, the authors in \cite{strasshofer2023} propose to find the parameters that maximize the \ac{BMI}, which is calculated by
\begin{equation}
    \MI(X;\Lapp) = 1 - \mathbb{E} \left[ \log_2 \left( 1 + \exp \left( -X \cdot \Lapp \right) \right) \right],
    \label{eq:BMI}
\end{equation}
where the expectation is realized via averaging over many frames in a Monte Carlo simulation.
Note that this framework works in a general setting and is not limited to the decoder analyzed in \cite{strasshofer2023}.

\section{Soft Information from a List and its Generating Algorithm}\label{sec:soft_info_from_list}
In the following, we first derive an approximate \emph{a posteriori} \ac{LLR} computation similar to~\cite{yuan2025a}.
We then demonstrate how this approach can be extended to \ac{BDD}-based list decoders.

\subsection{A Posteriori LLR Approximation}
Since the computation of the optimal a posteriori \acp{LLR} as defined in~\eqref{eq:lapp_def} is typically infeasible, an approximation with reduced computational complexity is required.

Given a list of candidate codewords $\List$, the $i$th a posteriori \ac{LLR} $\Lapp_i$ can be rewritten as
\begin{equation*}
    \Lapp_i = \ln\frac{\Pvcy(\Listi^0\mid\vy) + \Pvcy(\Codi^0\setminus\List\mid\vy) }{\Pvcy(\Listi^1\mid\vy) + \Pvcy(\Codi^1\setminus\List\mid\vy)},
\end{equation*}
where $\Listi^s = \List \cap \Codi^s \subseteq \Cod$ denotes the set of codewords in the list $\List$ that are $s\in\Ftwo$ at the $i$th position. 
The ratio of the likelihoods $\Pvcy(\Codi^s\setminus\List\mid\vy)$ can be approximated by the ratio of $\Pcy(s\mid y_i)$, following from
\begin{equation}\label{eq:ratio}
    \begin{aligned}
    \frac{\Pvcy(\Codi^0\setminus\List \mid \vy)}{\Pvcy(\Codi^1\setminus\List \mid \vy)}
    &= \frac{\Pcy(0\mid y_i)}{\Pcy(1\mid y_i)} \cdot 
    \frac{\Pvcy(\Codi^0\setminus\List \mid \vy_{[n]\setminus\{i\}})}{\Pvcy(\Codi^1\setminus\List \mid \vy_{[n]\setminus\{i\}})}  
    \\
    &\stackrel{\mathrm{(a)}}{\approx} \frac{\Pcy(0\mid y_i)}{\Pcy(1\mid y_i)}
    \cdot \frac{\vert\Codi^0\setminus\List\vert}{\vert\Codi^1\setminus\List\vert} 
    \stackrel{\mathrm{(b)}}{\approx} \frac{\Pcy(0\mid y_i)}{\Pcy(1\mid y_i)}.
    \end{aligned}
\end{equation}
For (a), we assume that the codewords outside of the list $\Codi^s \setminus \List$ sample the space $\Ftwo^n$ at coordinates $[n]\setminus\{i\}$ uniformly, and for (b), we consider the list $\List$ to be much smaller than $|\Codi^s|$.
As in \cite{yuan2025a}\footnote{Note that, unlike in \cite{yuan2025a}, where $\Pvcy(\Cod\setminus\List\mid\vy)$ denotes the probability that the transmitted codeword is not in the list, here it refers to the total probability of all the codeswords that are not in the list }, the individual likelihoods can then be found by \begin{equation*}
    \Pvcy(\Codi^s\setminus\List \mid \vy) \approx \Pvcy(\Cod\setminus\List\mid \vy) \cdot \Pcy(s\mid y_i).
\end{equation*}

Since it is infeasible to calculate $P(\Cod \setminus \List \mid \vy)$ directly, we introduce the covered space which allows to estimate it. %
\begin{definition}[Covered Space]\label{def:Vcov}
Given a code $\Cod$ and a list of codewords $\List\subseteq \Cod$, a covered space $\Vcov \subseteq \Ftwo^n$ fulfills
\begin{equation*}
    \List = \Vcov \cap \Cod.
\end{equation*}
\end{definition}
Consequently, any codeword not in the list is also not in the covered space, and thus $\Cod\setminus\List \subseteq\Ftwo^n\setminus\Vcov$.
Subsequently, we estimate the total likelihood of the codewords not found by the decoder as

\begin{equation*}
    \Pvcy(\Cod\setminus\List\mid \vy)\approx 2^{k-n} \cdot \left(1 - \Pvcy(\Vcov\mid\vy)\right),
\end{equation*}
following from
\begin{equation}\label{eq:phi}
    \frac{\Pvcy(\Cod\setminus\List\mid\vy)}{1-\Pvcy(\Vcov\mid\vy)}
    \stackrel{\mathrm{(a)}}{\approx} \frac{|\Cod\setminus\List|}{\left\lvert\Ftwo^n\setminus\Vcov\right\rvert}
    = \frac{2^k-|\List|}{2^n-|\Vcov|}
    \stackrel{\mathrm{(b)}}{\approx} 2^{k-n}.
\end{equation}
Similar to~\eqref{eq:ratio}, it is assumed for (a) that $\Cod \setminus \List$ samples the space $\Ftwo^n \setminus \Vcov$ uniformly, and for (b) that $|\List|$ and $|\Vcov|$ are negligibly small compared to $2^k$ and $2^n$, respectively.
Finally, for the $i$th a posteriori \ac{LLR}, the \acf{SOCS} is given by
\begin{equation}\label{eq:L_app_final}
    \Lapp_i \approx
    \ln\frac{
    \Pvcy(\Listi^0\mid\vy) 
    + 2^{k-n} \cdot (1-\Pvvy(\Vcov\mid \vy)) 
    \cdot \Pcy(0\mid y_i)
    }{
    \Pvcy(\Listi^1\mid\vy) 
    + 2^{k-n} \cdot (1-\Pvvy(\Vcov\mid \vy)) 
    \cdot \Pcy(1\mid y_i)
    },
\end{equation}
which is found equivalently\footnote{\hspace{1sp}\cite{yuan2025a} approximates~\eqref{eq:phi} as $(2^k-1)/(2^n-1)$. Evidently, the difference is negligible for practical $k$ and $n$.} in \cite{yuan2025a}.
To achieve a high accuracy, the decoder must aim to maximize $P(\Vcov \mid \vy)$.

\subsection{Covered Space of a List Decoder}
In \cite{yuan2025a}, the covered space $\Vcov$ for \eqref{eq:L_app_final} is chosen by \ac{GRAND}.
Once the desired number of candidate codewords $|\List|$ is found, 
the likelihood of the covered space 
$\Pvvy(\Vcov\mid\vy)$ is determined as the sum of the probabilities ${\Pvvy(\vchd \oplus \vn\mid \vy)}$ for all noise queries~$\vn$.

Unlike \ac{GRAND}, which inherently discovers non-codeword vectors $\vv \in \Ftwo^n\setminus \Cod$ that have high probability $\Pvvy(\vv\mid \vy)$, general list decoders only determine high-probability codewords.
The following proposition shows how \ac{BDD}-based list decoders, such as the Chase-II decoder, can use their inherent capabilities to consider high-pro\-ba\-bi\-li\-ty non-code\-word vectors for the generation of soft information.

\begin{proposition}
[Covered Space of a \ac{BDD}-based List Decoder]\label{prop:V_LD}
Let $\List$ denote the set of codewords obtained by decoding a code $\Cod$ with minimum distance $\dmin$, using \ac{BDD} with error-correcting capability $t$ applied to a set of testwords $\Test$.
Then, a covered space of the list decoder is given by
\begin{equation}
    \Vld = \Ball_{\dmin - 1}(\List) \cup \Ball_{t}(\Test).
    \label{eq:vcov_def}
\end{equation}
\end{proposition}

\begin{IEEEproof}
Since $\Cod$ has minimum distance $\dmin$, the Hamming ball $\Ball_{\dmin - 1}(\vc)$ around a codeword $\vc \in \Cod$ contains no other codeword.
Therefore, the union of the Hamming balls around the codewords in $\List$ satisfies
\begin{equation}\label{eq:list_balls_Vcov}
    \Ball_{\dmin - 1}(\List) \cap \Cod = \List.
\end{equation}
When \ac{BDD} with error-correcting capability 
$t$
is applied to a testword $\ttest \in \Test$, a codeword $\vc \in \Cod$ is obtained if and only if $\vc \in \Ball_t(\ttest)$. 
Hence, we find that
\begin{equation}\label{eq:test_balls_Vcov}
    \Ball_t(\Test) \cap \Cod = \List.
\end{equation}
From~\eqref{eq:list_balls_Vcov} and~\eqref{eq:test_balls_Vcov}, it follows that both sets $\Ball_{\dmin - 1}(\List) $ and $ \Ball_{t} (\Test)$ are a covered space according to~\cref{def:Vcov}.
Consequently, $\Vld$ is also a covered space as 
\begin{equation*}
    \Vld \cap \Cod = \left( \Ball_{\dmin - 1}(\List) \cup \Ball_t(\Test) \right) \cap \Cod = \List.
\end{equation*}
\end{IEEEproof}

\emph{Remark}: In the case of $\lvert\Test\rvert=0$, \cref{prop:V_LD} applies to any list decoder.
The approach of \ac{SOGRAND} from \cite{yuan2025a} is reflected by $\Vcov=\Ball_0(\Test)=\Test$.

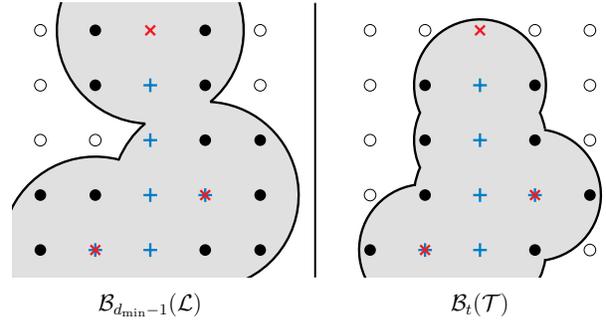
\begin{figure}
    \centering
    \resizebox{1.\linewidth}{!}{\begin{tikzpicture}

\def\centerarc[#1](#2)(#3:#4:#5)%
    { \draw[#1] ($(#2)+({#5*cos(#3)},{#5*sin(#3)})$) arc (#3:#4:#5); }
    
\tikzset{cross/.style={cross out, draw, 
     minimum size=5pt, 
     very thick,
     inner sep=0pt, outer sep=0pt}}

\def\radius{1}
\def\radiusouter{1.4142}
\def\radiuslist{1.7}
\def\radiuschase{1.2}
\def\dotsize{3pt}
\def\circlesize{2.5pt}

\draw[black, line width=1pt] (5,-0.5) -- (5,4.5);
\node[] at (2,-1) {\large $\Ball_{\dmin - 1}(\List)$};
\node[] at (8,-1) {\large $\Ball_{t}(\Test)$};

\begin{scope}
\clip(-0.5, -0.5) rectangle (5.5,4.5);

\foreach \x in {0,1,2,3,4} {
    \foreach \y in {0, 1, 2, 3, 4, 5}
        \draw[black] (\x,\y) circle (\dotsize);

}

\foreach \x/\y in {3/1, 2/4, 1/0} {
    \fill[fill=hellgrau] (\x, \y) circle (\radiuslist);

    \foreach \angle in {0, 90, 180, 270} {
        \fill[black] ({\x + \radius*cos(\angle)}, {\radius*sin(\angle)+\y}) circle (\dotsize);
    }
    \foreach \angle in {45, 135, 225, 315} {
        \fill[black] ({\x + \radiusouter*cos(\angle)}, {\radiusouter*sin(\angle) + \y}) circle (\dotsize);
    }
}

\foreach \x/\y in {3/1, 2/4, 1/0} {
    \draw (\x,\y) node[cross, color=rot] {};
}

\foreach \x/\y in {2/0, 2/1, 2/2, 2/3, 3/1, 1/0} {
    \fill[hellgrau] (\x,\y) circle ({\dotsize*2});
    \draw (\x,\y) node[cross, color=mittelblau, rotate=45] {};
}

\foreach \x/\y in {3/1, 1/0} {
    \fill[hellgrau] (\x,\y) circle ({\dotsize*2});
    \draw (\x,\y) node[cross, color=mittelblau, rotate=45] {};
    \draw (\x,\y) node[cross, color=rot] {};
}

\centerarc[black,very thick](3,1)(-75:87:\radiuslist)
\centerarc[black,very thick](2,4)(-50:267:\radiuslist)
\centerarc[black,very thick](3,1)(130:158:\radiuslist)
\centerarc[black,very thick](1,0)(75:160:\radiuslist)

\end{scope}

\begin{scope}[shift={(6,0)}]

\clip(-0.5, -0.5) rectangle (5.5,4.5);

\foreach \x in {0,1,2,3,4} {
    \foreach \y in {0, 1, 2, 3, 4, 5}
        \draw[black] (\x,\y) circle (\dotsize);
}

\foreach \x/\y in {2/0, 2/1, 2/2, 2/3, 3/1, 1/0} {
    \fill[fill=hellgrau] (\x, \y) circle (\radiuschase);

    \foreach \angle in {0, 90, 180, 270} {
        \fill[black] ({\x + \radius*cos(\angle)}, {\radius*sin(\angle)+\y}) circle (\dotsize);
    }
}
\foreach \x/\y in {3/1, 2/4, 1/0} {
    \fill[hellgrau] (\x,\y) circle ({\dotsize*2});
    \draw (\x,\y) node[cross, color=rot] {};
}

\foreach \x/\y in {2/0, 2/1, 2/2, 2/3, 3/1, 1/0} {
    \fill[hellgrau] (\x,\y) circle ({\dotsize*2});
    \draw (\x,\y) node[cross, color=mittelblau, rotate=45] {};
}

\foreach \x/\y in {3/1, 1/0} {
    \fill[hellgrau] (\x,\y) circle ({\dotsize*2});
    \draw (\x,\y) node[cross, color=mittelblau, rotate=45] {};
    \draw (\x,\y) node[cross, color=rot] {};
}

\centerarc[black,very thick](2,0)(-75:-8:\radiuschase)
\centerarc[black,very thick](3,1)(-82:82:\radiuschase)
\centerarc[black,very thick](2,2)(10:25:\radiuschase)
\centerarc[black,very thick](2,3)(-25:205:\radiuschase)
\centerarc[black,very thick](2,2)(155:205:\radiuschase)
\centerarc[black,very thick](2,1)(155:172:\radiuschase)
\centerarc[black,very thick](1,0)(98:230:\radiuschase)

\end{scope}

\end{tikzpicture}}
    \caption{ A portion of $\Ftwo^n$ is visualized, where a list decoder has found the codewords in $\List$ (marked by \protect\redcross{}), from decoding the testwords in  $\Test$ (marked by \protect\bluecross{}). The vectors in the space covered by $\Ball_{\dmin -1}(\List)$ and $\Ball_{t}(\Test)$ are depicted by \protect\blackdot{}. Vectors outside of the Hamming balls are indicated by \protect\greydot.}
    \label{fig:covered_space}
\end{figure}

The sets $\Ball_{\dmin - 1}(\List) $ and $\Ball_{t} (\Test)$ are visualized in \cref{fig:covered_space}.
Calculating the union in \eqref{eq:vcov_def} and the corresponding probability $\Pvvy(\Vld\mid \vy)$ is computationally complex, particularly due to the large overlap of the Hamming balls.
Therefore, we introduce efficient approximations for $\Pvvy(\Vld \mid \vy)$ in the following section.

\section{Calculating the Probability of the Covered Space} \label{sec:calulating_vcov}
To estimate the probability $\Pvvy(\Vld \mid \vy)$ of the space covered by a list decoder, we propose two methods.  
The first, $\Pvvy(\Vld \mid \vy) \approx P(\Ball_{r}(\List) \mid \vy)$, is applicable to any list decoder.  
The second, $\Pvvy(\Vld \mid \vy) \approx P(\Ball_{t}(\Test) \mid \vy)$, is specifically tailored to the Chase-II decoding algorithm.  
Both methods use the following simple approach for calculating the probability of all vectors within a Hamming ball.

Consider a vector $\vv\in \Ftwo^n$ with probability $\Pvvy(\vv\mid\vy)$ as given in~\eqref{eq:cwpros_from_gamma}.
Observe that for all vectors around $\vv$, only one position needs to be adjusted to find the probability.
Thus, the total probability of all vectors in the Hamming ball $\Ball_{1}(\vv)$ can be computed with 
\begin{equation*}
    \begin{aligned}
        &\Pvvy\left(\Ball_{1}(\vv)\mid \vy\right) = \Pvvy(\vv\mid\vy) + \sum\limits_{i \in [n]} \Pvvy(\vv\oplus \one_i\mid\vy)\\ 
        &= \Pvvy(\vv\mid\vy)
        \cdot \underbrace{\bigg( 1 +\sum\limits_{i:\,v_i=\chd_i} \frac{1-\gamma_i}{\gamma_i} + \sum\limits_{i:\,v_i\neq\chd_i}  \frac{\gamma_i}{1-\gamma_i} \bigg)}_{\triangleq\,b_1(\vv,\vy)},
    \end{aligned}
\end{equation*}
where $b_1(\vv, \vy)$ is the \emph{Hamming ball factor} corresponding to the Hamming ball $\Ball_{1}(\vv)$.

\begin{proposition}[Hamming ball factor]\label{prop:ball_factor}
Given a vector $\vv\in\Ftwo^n$, the Hamming ball factor $b_r(\vv, \vy)$ is found such that %
\begin{equation*}
    \Pvvy(\Ball_{r}(\vv)\mid \vy)\\ = b_r(\vv,\vy) \cdot \Pvvy(\vv\mid\vy).
\end{equation*}
It can be calculated using the Hamming ball summand
\begin{equation*}
    s_r(\vv,\vy) = \sum\limits_{i \in [n]} \exp \left( - r \cdot \left\lvert \Lin_i \right\rvert \cdot (-1)^{v_i \oplus \chd_i} \right).
\end{equation*} 
For Hamming balls with radius $r=1$, it holds that
\begin{equation*}
    b_1(\vv,\vy) = 1 + s_1(\vv,\vy),
\end{equation*}
and for radius $r=2$, the multiplier can be derived as
\begin{equation*}
    b_2(\vv,\vy)= b_1(\vv,\vy) + \frac{1}{2} \cdot \left( s_1(\vv,\vy)^2
    - s_2(\vv,\vy) \right).
\end{equation*}

\end{proposition}

\subsection{Union of List Candidate Hamming Balls}
The probability of the covered space $P(\Vld\mid\vy)$ as defined in~\eqref{eq:vcov_def} can be approximated as $P(\Vld\mid\vy) \approx P(\Ball_r (\List)\mid\vy)$.
By choosing ${r\leq \lfloor\frac{\dmin-1}{2}\rfloor}$, the union between the individual Hamming balls has no overlap, and thus
\begin{equation*}
    \Pvcy ( \Ball_{r}(\List) \mid \vy) = \sum\limits_{\vc \in \List} \Pvcy ( \Ball_{r}(\vc) \mid \vy )= \sum\limits_{\vc \in \List} b_r(\vc,\vy) \cdot \Pvcy(\vc \mid \vy) ,
\end{equation*}
where $b_r(\vc,\vy)$ is the Hamming ball factor as introduced in \cref{prop:ball_factor}.

\subsection{Union of Chase Pattern Hamming Balls}
Similarly, we can estimate the probability of the covered space as $P(\Vld\mid\vy) \approx P(\Ball_r(\Test)\mid\vy)$.
For the Chase-II algorithm, the testwords $\ttest\in\Test$ only differ in the $p=|\Ind|$ least reliable positions $\Ind$.
All other positions $[n]\setminus \Ind$ are identical to the hard decision as $\ttest_{[n]\setminus \Ind} = \vchd_{[n]\setminus \Ind}$ for all testwords.
Therefore, we can calculate $\Pvvy(\Test \mid \vy)$ as follows

\begin{equation}
\begin{aligned}
    & \Pvvy(\Test \mid \vy) = \sum\limits_{\Kcal\in \pws_{\Ind}} \Pvcy\left(\vv = \vchd \oplus \one_{\Kcal} \mid \vy\right)\\
    &= \sum\limits_{\Kcal\in \pws_{\Ind}} \prod\limits_{i\in [n]\setminus \Ind} \gamma_i 
    \cdot\prod\limits_{i\in \Ind \setminus \Kcal} \gamma_i 
    \cdot\prod\limits_{i\in \Kcal} (1-\gamma_i)\\
    &= \prod\limits_{i\in [n]\setminus \Ind} \gamma_i \cdot \underbrace{ \sum\limits_{\Kcal\in \pws_{\Ind}} \;
    \prod\limits_{i\in \Ind \setminus \Kcal} \gamma_i 
    \cdot\prod\limits_{i\in \Kcal} (1-\gamma_i) }_{=1}.
\end{aligned}\label{eq:chase_balls}
\end{equation}

Hence, $\Pvvy(\Test \mid \vy)$ can be calculated as the product of the likelihoods that the hard decisions at the positions ${[n]\setminus \Ind}$ are correct. 
Since the Chase-II algorithm covers all possibilities for the positions $\Ind$, the sum of the likelihoods is $1$.

Following from the uniqueness of each testword at positions $\Ind$, observe that the $n-p$ dimensional Hamming balls in the coordinates $[n]\setminus \Ind$, i.e., $\Ball_r(\ttest, \Ind)$, do not overlap.
Combining this with \eqref{eq:chase_balls}, we obtain
\begin{equation*}
    \Pvvy(\Ball_r(\Test) \mid \vy) = b_r\left(\vchd,\vy, \Ind \right) \cdot \prod\limits_{j\in [n]\setminus \Ind} \gamma_j,
\end{equation*} %
where $b_r\left(\vchd,\vy, \Ind \right)$ is the Hamming ball factor with respect to the coordinates $[n]\setminus\Ind$. Note that the Hamming ball factors $b_r\left(\vchd,\vy, \Ind \right)$ for $r\in\{1,2\}$ can be obtained by replacing the set $[n]$ with the set of considered coordinates $[n]\setminus\Ind$ in the calculations according to \cref{prop:ball_factor}.

We now turn to \ac{SOCS} decoding which uses the above-de\-scri\-bed estimations of the probability $P(\Vld\mid\vy)$ of the covered space.

\section{Turbo Product Decoding Using SOCS}
For the decoding of \acp{PC} using \ac{SOCS}, the a posteriori \ac{LLR} is calculated according to \eqref{eq:L_app_final}. The probabilities of the codewords in the list $\Pvcy(\List_i^0 \mid \vy)$ and $\Pvcy(\List_i^1 \mid \vy)$ are calculated by \eqref{eq:cwpros_from_gamma} while the probability of the covered space $P(\Vld\mid\vy)$ can be found as shown in \cref{sec:calulating_vcov}.
We denote the \ac{SOCS} decoders by $\ac{SOCS}(\Vcov)$, where $\Vcov$ is the respective covered space used.
For comparison, an alternative decoder, denoted by $\ac{SOCS}(\beta)$, is proposed that approximates the likelihood of the codewords outside of the list as constant, i.e., $\Pvcy(\Cod\setminus\List\mid \vy) \approx \beta$.
The \ac{LLR} passed between row and column decoders is found via
\begin{equation}\label{eq:socs_lext}
    \Lext_i = \alpha \cdot \left( \Lapp_i - \Lin_i \right),
\end{equation}
where $\alpha$ is a scaling parameter that is introduced to improve the decoding performance accounting for the suboptimality of the LLR computation and the iterative decoding algorithm in general.

\section{Results}\label{sec:results}

\begin{figure*}[htp]
	\centering
	\begin{tikzpicture}
\begin{groupplot}[
    group style={
        group size= 2 by 1, 
        horizontal sep=1.45cm,
    },
    ymode = log,
    height=\columnwidth,
	xmajorgrids,
	xminorgrids=true,
	ymajorgrids,
	yminorticks=true,
	major grid style={hellgrau},
    minor grid style={densely dotted, hellgrau},
	yminorgrids=true,
    legend pos=north east,
	legend style={fill,fill opacity=0.75, text opacity=1, draw=none},
    tick align=outside,
    tick pos=left,
    xlabel={$E_\mathrm{b}/N_0$ in dB},
    xtick={3.5,4,...,5.5},
    xminorgrids=true,
    minor x tick num=1,
    xmin=3.5,
    xmax=5.5,
    xtick style={color=black},
    ymajorgrids,
    ymin=1E-8, 
    ymax=1E-1, 
    ytick style={color=black},
    label style = {font=\small},
    ticklabel style = {font=\small},
    every axis plot/.style={
        line width=0.65pt,
        mark size=2.2pt
    }
]
\nextgroupplot[
    title={(a) $(256,247,4)^2$, $t=1$},
    xtick={3.5,4,...,5.5},
    xmin=3.5,
    xmax=5,
    ylabel={BER},
    width=\columnwidth,
]

\addplot [black, dotted, postaction={decorate,decoration={raise=2pt,text align={left, left indent=8.47cm},text along path,text={|\scriptsize| HD rate-distortion bound}}}, forget plot] 
table[col sep=comma]{
3.50000000, 1.008e-02
3.61111111, 8.935e-03
3.72222222, 7.849e-03
3.83333333, 6.821e-03
3.94444444, 5.851e-03
4.05555556, 4.938e-03
4.16666667, 4.082e-03
4.27777778, 3.284e-03
4.38888889, 2.543e-03
4.50000000, 1.861e-03
4.56000000, 1.518e-03
4.62000000, 1.193e-03
4.68000000, 8.867e-04
4.74000000, 6.013e-04
4.80000000, 3.390e-04
4.81000000, 2.979e-04
4.83000000, 2.182e-04
4.84000000, 1.798e-04
4.85700000, 1.170e-04
4.86400000, 9.226e-05
4.87100000, 6.827e-05
4.88500000, 2.333e-05
4.88839400, 1.335e-05
4.89009100, 8.588e-06
4.89340800, 1.491e-07
4.89348500, 1.238e-09
};
\label{plot:hd_rate_dis_bound}

\addplot [black, solid, postaction={decorate,decoration={raise=2pt,text align={left, left indent=2.805cm},text along path,text={|\scriptsize| SD rate-distortion bound}}}, forget plot] 
table[col sep=comma]{
3.50000000, 9.524e-04
3.52222222, 8.511e-04
3.54444444, 7.523e-04
3.56666667, 6.560e-04
3.58888889, 5.623e-04
3.61111111, 4.712e-04
3.63333333, 3.832e-04
3.65555556, 2.982e-04
3.67777778, 2.168e-04
3.70000000, 1.395e-04
3.70512222, 1.224e-04
3.71024444, 1.055e-04
3.71536667, 8.900e-05
3.72048889, 7.283e-05
3.72561111, 5.706e-05
3.73073333, 4.175e-05
3.73585556, 2.702e-05
3.74097778, 1.306e-05
3.74610000, 5.828e-07
3.74618571, 4.090e-07
3.74634490, 1.016e-07
3.74635714, 7.935e-08
3.74637551, 4.675e-08
3.74638776, 2.572e-08
3.74639388, 1.554e-08
3.74640000, 5.742e-09
};
\label{plot:sd_rate_dis_bound}

\addplot[brown, dashdotted, mark=asterisk, mark options={solid}]
table[col sep=comma]{
3.50,9.34793E-02
3.60,9.21597E-02
3.70,9.06839E-02
3.80,8.86829E-02
3.90,8.64886E-02
4.00,8.33241E-02
4.10,7.93588E-02
4.20,7.33106E-02
4.30,6.01687E-02
4.40,2.48571E-02
4.50,2.35996E-03
4.60,3.55063E-05
4.70,2.20809E-07
4.80,1.07886E-08
};
\label{plot:CP}

\addplot[apfelgruen, dashdotted, mark=triangle, mark options={solid}]
table[col sep=comma]{
3.5, 0.029435,
3.6, 0.0276446,
3.7, 0.0255973,
3.8, 0.0234726,
3.9, 0.0210362,
4, 0.0178951,
4.1, 0.0148171,
4.2, 0.0102555,
4.3, 0.00498471,
4.4, 0.00102735,
4.5, 4.2056e-05,
4.6, 7.3939e-07,
4.7, 3.9367e-08,
4.8, 7.42355e-09,
};
\label{plot:CP_opt}

\addplot[orange, dashed, mark=diamond, mark options={solid}]
table[col sep=comma]{
3.5, 0.0241345
3.6, 0.0225601
3.7, 0.020537
3.8, 0.0184345
3.9, 0.0158923
4, 0.0127264
4.1, 0.00948476
4.2, 0.00471213
4.3, 0.001393
4.4, 0.000125755
4.5, 2.59434e-06
4.6, 2.60027e-08
4.65, 2.7573e-09
};
\label{plot:SOCCP}

\addplot [cyan, dashed, mark=+,  mark options={solid}] 
table[col sep=comma]{
3.5, 0.0231226
3.6, 0.0214245
3.7, 0.0197891
3.8, 0.0177746
3.9, 0.0156236
4, 0.0127834
4.1, 0.00821008
4.2, 0.00331927
4.3, 0.0003503
4.4, 7.90666e-06
4.5, 2.21231e-08
4.55, 3.20759e-09
};
\label{plot:SOCS_lb}

\addplot [violettblau, dashed, mark=x,  mark options={solid}] 
table[col sep=comma]{
3.5, 0.0227615
3.6, 0.0211301
3.7, 0.019191
3.8, 0.0172002
3.9, 0.014754
4, 0.0117236
4.1, 0.00776829
4.2, 0.00239472
4.3, 0.000150292
4.4, 2.32657e-06
4.5, 4.31736e-09
};
\label{plot:SOCS}

\addplot[magenta, dashed, mark=o, mark options={solid}]
table[col sep=comma]{
3.5, 0.0229304
3.6, 0.0212417
3.7, 0.0194214
3.8, 0.0175456
3.9, 0.0150788
4, 0.0117943
4.1, 0.0077372
4.2, 0.00246592
4.3, 0.000158821
4.4, 2.31344e-06
4.5, 4.24787e-09
};
\label{plot:SOCS_nb}

\addplot[rot, dashdotted, mark=star,]
table[col sep=comma]{
2, 0.0429148
2.1, 0.0411531
2.2, 0.0392876
2.3, 0.0376316
2.4, 0.0360807
2.5, 0.034499
2.6, 0.0328402
2.7, 0.0312435
2.8, 0.0298587
2.9, 0.028274
3.0, 0.0269143
3.1, 0.0254745
3.2, 0.0240163
3.3, 0.0225047
3.4, 0.0210476
3.5, 0.0195679
3.6, 0.0181063
3.7, 0.016325
3.8, 0.0143513
3.9, 0.0119327
4.0, 0.00829593
4.1, 0.00308771
4.2, 0.000203135
4.3, 1.17271e-06
4.4, 2.81092e-09
};
\label{plot:MAP}

\nextgroupplot[
    title={(b) $(256,239,6)^2$, $t=2$},
    xtick={3.2,3.4,...,4.2},
    xmin=3.3,
    xmax=4.1,
    ylabel={\color{white}BER},
    width=\columnwidth,
]

\draw[-{Triangle[]}] (axis cs: 3.365, 3e-7) -- (axis cs: 3.305, 3e-7);
\node[rotate=-90 ]  at (axis cs: 3.38, 3e-7){\scriptsize SD rate-distortion bound};

\addplot [black, dotted, postaction={decorate,decoration={raise=2pt,text align={left, left indent=8.815cm},text along path,text={|\scriptsize| HD rate-distortion bound}}}, forget plot] 
table[col sep=comma]{
3.25000000, 7.107e-03
3.35000000, 5.978e-03
3.45000000, 4.908e-03
3.55000000, 3.899e-03
3.65000000, 2.953e-03
3.75000000, 2.074e-03
3.81000000, 1.581e-03
3.87000000, 1.118e-03
3.90000000, 8.977e-04
3.92000000, 7.560e-04
3.94000000, 6.185e-04
3.96000000, 4.857e-04
3.98000000, 3.581e-04
4.00600000, 2.016e-04
4.01200000, 1.673e-04
4.02400000, 1.013e-04
4.03000000, 7.000e-05
4.03540000, 4.305e-05
4.03720000, 3.440e-05
4.03900000, 2.596e-05
4.04000000, 2.138e-05
4.04200000, 1.251e-05
4.04300000, 8.268e-06
4.04400000, 4.208e-06
4.04446212, 2.425e-06
4.04469318, 1.568e-06
4.04492424, 7.454e-07
4.04515530, 2.016e-10
};
\label{plot:hd_rate_dis_bound2}

\addplot[brown, dashdotted, mark=asterisk, mark options={solid}]
table[col sep=comma]{
3.2, 0.0451101,
3.25, 0.0433758,
3.3, 0.041559,
3.35, 0.0393377,
3.4, 0.0370165,
3.45, 0.0345993,
3.5, 0.0316327,
3.55, 0.0285466,
3.6, 0.0244035,
3.65, 0.0211276,
3.7, 0.0146736,
3.75, 0.00894036,
3.8, 0.00424957,
3.85, 0.00101013,
3.9, 0.000162306,
3.95, 7.60422e-06,
4, 2.42247e-07,
4.05, 3.10509e-09,
};

\addplot[apfelgruen, dashdotted, mark=triangle, mark options={solid}]
table[col sep=comma]{
3.2, 0.0356698,
3.25, 0.0341369,
3.3, 0.0327455,
3.35, 0.031025,
3.4, 0.0290625,
3.45, 0.0266187,
3.5, 0.0239066,
3.55, 0.021374,
3.6, 0.0175655,
3.65, 0.0134541,
3.7, 0.00931178,
3.75, 0.00420163,
3.8, 0.00163132,
3.85, 0.00031777,
3.9, 2.96631e-05,
3.95, 1.37176e-06,
4, 4.76914e-08,
4.05, 2.3142e-09,
};

\addplot[orange, dashed, mark=diamond, mark options={solid}]
table[col sep=comma]{
3.2, 0.0352689,
3.25, 0.0336481,
3.3, 0.031716,
3.35, 0.0297361,
3.4, 0.0272359,
3.45, 0.0237347,
3.5, 0.0198401,
3.55, 0.0150232,
3.6, 0.00885487,
3.65, 0.00279135,
3.7, 0.000517684,
3.75, 3.69297e-05,
3.8, 5.1745e-07,
3.85, 7.77789e-09,
};

\addplot [cyan, dashed, mark=+,  mark options={solid}] 
table[col sep=comma]{
3.2, 0.0314112,
3.25, 0.0299776,
3.3, 0.0279623,
3.35, 0.0258143,
3.4, 0.0229701,
3.45, 0.019905,
3.5, 0.0150976,
3.55, 0.00930376,
3.6, 0.00305519,
3.65, 0.00065627,
3.7, 4.46771e-05,
3.75, 1.64078e-06,
3.8, 2.64396e-08,
3.82, 4.02393e-09,
};

\addplot [violettblau, dashed, mark=x,  mark options={solid}] 
table[col sep=comma]{
3.2, 0.0312584,
3.25, 0.0297502,
3.3, 0.0276917,
3.35, 0.0253061,
3.4, 0.0226812,
3.45, 0.0182898,
3.5, 0.0115871,
3.55, 0.00533816,
3.6, 0.00124594,
3.65, 0.000152536,
3.7, 6.26324e-06,
3.75, 7.45276e-08,
3.775, 6.23481e-09,
};

\addplot[magenta, dashed, mark=o, mark options={solid}]
table[col sep=comma]{
3.2, 0.0316986,
3.25, 0.0299669,
3.3, 0.0279344,
3.35, 0.0259309,
3.4, 0.0231628,
3.45, 0.019774,
3.5, 0.0133521,
3.55, 0.00855046,
3.6, 0.00285647,
3.65, 0.000482127,
3.7, 4.4195e-05,
3.75, 1.07919e-06,
3.81, 4.28282e-09,
};

\end{groupplot}

\renewcommand{\arraystretch}{1.15}

\node (tab) [
    below = 0.1cm of current bounding box.south,
    shape=rectangle,
    draw=none,
    anchor=north,
    font=\footnotesize
] {
\begin{NiceTabular}{l|*{4}{c}}
    \toprule
    Baseline 
    & \multicolumn{4}{c}{
        \begin{tabular}{@{}ccc@{}}
        \ref{plot:CP} CP \cite{pyndiah1998} 
        & \ref{plot:CP_opt} Optimized CP \cite{strasshofer2023}  
        & \ref{plot:MAP} Iterative MAP \\
        \end{tabular}
    } \\
    \midrule
    Proposed 
        & \ref{plot:SOCCP} $\ac{SOCS}(\beta)$ 
        & \ref{plot:SOCS_lb} $\ac{SOCS}\left(\Ball_t(\List)\right)$
        & \ref{plot:SOCS} $\ac{SOCS}\left(\Ball_t(\Test)\right)$ 
        & \ref{plot:SOCS_nb} $\ac{SOCS}\left(\Test\right)$
        \\
    \bottomrule
\end{NiceTabular}
};

\end{tikzpicture}
	\vspace{-0.1cm}
	\caption{\footnotesize \acs{BER} comparison of an extended Hamming \ac{PC} (a) and an extended BCH \ac{PC} (b) with $4$ iterations and $p=5$ for Chase decoding.}
	\label{fig:Ham_BCH_256}
\end{figure*}
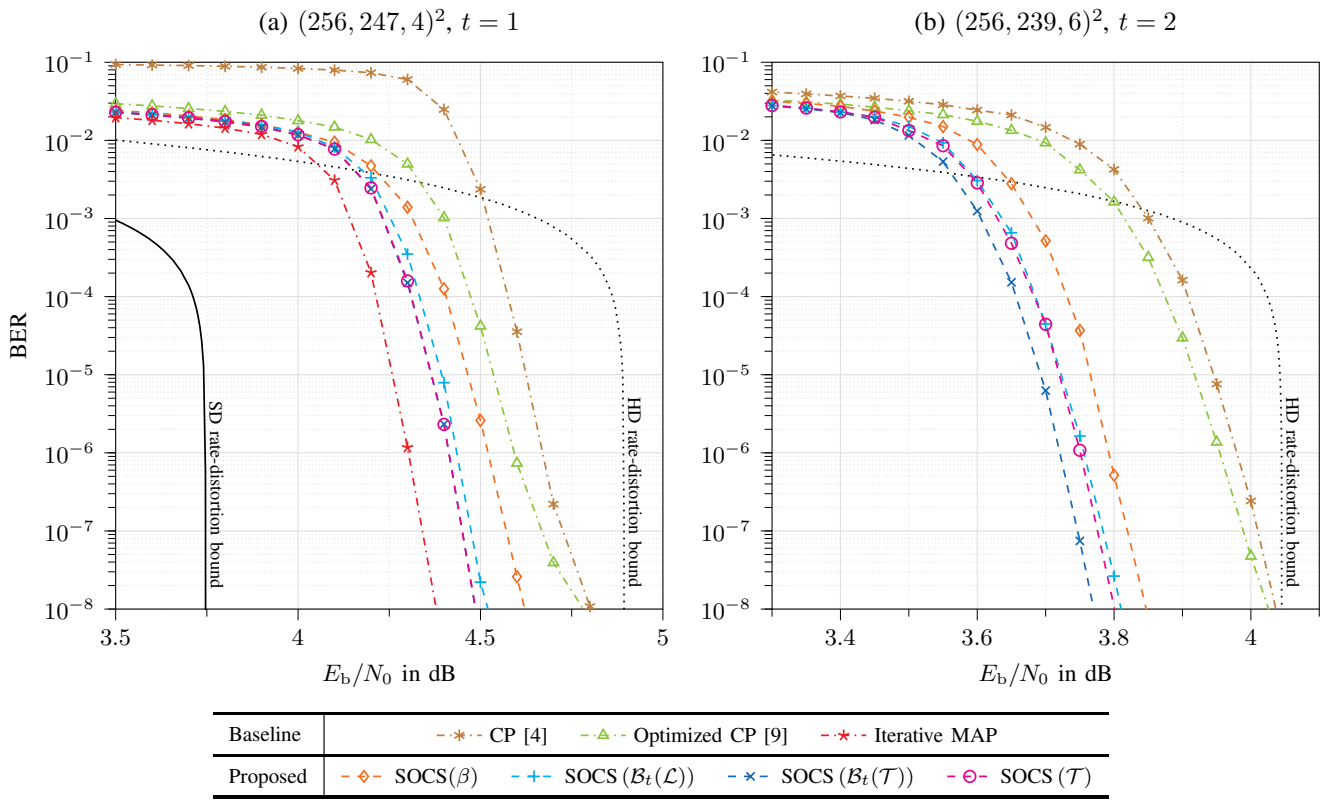

The simulation results for \acp{PC} with $(256,247,4)$ extended Hamming component codes and $(256,239,6)$ extended \ac{BCH} component codes are depicted in \cref{fig:Ham_BCH_256}a and \cref{fig:Ham_BCH_256}b, respectively.

As baselines, the \ac{HD} and \ac{SD} rate-distortion bounds, as well as the \ac{BER} curves for \ac{CP} decoding as described in \cite{pyndiah1998} and optimized \ac{CP} decoding as proposed in \cite{strasshofer2023} are given. 
For the extended Hamming \ac{PC}, iterative MAP labels the \acp{BER} obtained from iterative optimal a posteriori decoding as in \cite{goalic2002}, which uses the decoder from \cite{ashikhmin2004}\hspace{1sp}\cite{nazarov1998}.

To have a fair comparison, all the list-based decoders generate the list using the Chase-II algorithm with $p=5$. In the last half iteration, the estimated component codewords are found by choosing the most likely codeword in the respective lists generated by Chase-II decoding.
The scaling parameters $\alpha$, and, if applicable, $\beta$ for the \ac{SOCS} and optimized \ac{CP} decoders are found by a grid search maximizing the \ac{BMI} according to \eqref{eq:BMI} for each half iteration with soft-output at roughly the same \ac{BER}. The optimized values can be found in \cref{tab:parameters_ham} and \cref{tab:parameters_bch}.

For extended Hamming \acp{PC}, the \ac{SOCS} decoders can reduce the gap to iterative optimal a posteriori decoding at a $\ac{BER}$ of $10^{-6}$ to $0.1 \, \mathrm{dB}$. 
The \ac{SOCS} decoders with estimated $P(\Vcov \mid \vy)$ gain roughly $ 0.1 \, \mathrm{dB}$ compared to the $\ac{SOCS}(\beta)$ decoder
while $\ac{SOCS}(\beta)$ in turn outperforms optimized and classic \ac{CP} decoding.

The small gap between iterative optimal a posteriori and $\ac{SOCS}(\Vcov)$ implies that the extrinsic \ac{LLR} is closely approximated. 
$\ac{SOCS}(\Ball_1(\Test))$ and $\ac{SOCS}(\Test)$ performing the same implies that $P(\Ball_1(\Test)\mid \vy) \approx P(\Test \mid \vy)$, which can be explained by the small radius of the Hamming balls.

For extended BCH \acp{PC}, the \ac{SOCS} decoders gain up to $0.25\,\mathrm{dB}$ compared to \ac{CP} decoding.
With $t=2$, $\ac{SOCS}(\Ball_2(\Test))$ covers a much larger space than $\ac{SOCS}(\Test)$, significantly improving the estimated a posteriori \ac{LLR} and decreasing the \ac{BER}.
With the larger covered space, $\ac{SOCS}(\Ball_2(\List))$ and $\ac{SOCS}(\Test)$ perform similarly.

Note that $\ac{SOCS}(\Ball_t(\Test))$ provides a smaller \ac{BER} than the $\ac{SOCS}(\Ball_t(\List))$ indicating that the soft information obtained by using the testwords to obtain $P(\Vld\mid\vy)$ is better than that of the codewords in the list. 
This can be explained by the fact that there are more testwords than codewords in the list, and that the testwords typically have higher likelihoods.
Thus, the estimation of $P(\Cod \setminus \List \mid \vy)$ is more accurate.

Approximating $P(\Cod \setminus \List \mid \vy)$ as constant, as for $\ac{SOCS}(\beta)$, reduces the performance compared to the other \ac{SOCS} decoders. 
However, compared to the optimized \ac{CP} decoder, it performs better. 
This improvement stems from the case when no competitor is found, i.e., all codewords in the list agree in the corresponding position. 
Then, $\ac{SOCS}(\beta)$ produces an output dependent on $P(\List\mid\vy)$, while the optimized \ac{CP} decoder returns a constant output. Therefore, the a posteriori \ac{LLR} estimate of $\ac{SOCS}(\beta)$ is more accurate than that of \ac{CP}.

\section{Conclusion}\label{sec:conclusion}
In this paper, we proposed a new soft-output decoder, called \ac{SOCS} decoder, that approximates the a posteriori \ac{LLR} for any list decoder using its covered space.

For \ac{TPD}, we show that \ac{SOCS} decoding gains up to $0.25\,\mathrm{dB}$ compared to \ac{CP} decoding.
The simulation results indicate that the accuracy of the reliability found by \ac{CP} at positions without competitors can be improved.
However, as \ac{SOCS} decoding requires calculations in the probability domain, it is not yet competitive with \ac{CP} decoding in terms of practical implementations.
Nevertheless, \ac{SOCS} decoding provides valuable insights into how to improve soft-output decoding.

As a result, future work aimed at improving \ac{CP} decoding should focus on finding a better reliability estimate for positions without competing codewords.

\bibliographystyle{IEEEtran}
\bibliography{bibliofile}

\begin{thebibliography}{10}
\providecommand{\url}[1]{#1}
\csname url@samestyle\endcsname
\providecommand{\newblock}{\relax}
\providecommand{\bibinfo}[2]{#2}
\providecommand{\BIBentrySTDinterwordspacing}{\spaceskip=0pt\relax}
\providecommand{\BIBentryALTinterwordstretchfactor}{4}
\providecommand{\BIBentryALTinterwordspacing}{\spaceskip=\fontdimen2\font plus
\BIBentryALTinterwordstretchfactor\fontdimen3\font minus
  \fontdimen4\font\relax}
\providecommand{\BIBforeignlanguage}[2]{{%
\expandafter\ifx\csname l@#1\endcsname\relax
\typeout{** WARNING: IEEEtran.bst: No hyphenation pattern has been}%
\typeout{** loaded for the language `#1'. Using the pattern for}%
\typeout{** the default language instead.}%
\else
\language=\csname l@#1\endcsname
\fi
#2}}
\providecommand{\BIBdecl}{\relax}
\BIBdecl

\bibitem{elias1954}
P.~Elias, ``Error-free {{Coding}},'' \emph{Trans. IRE Prof. Group Inf. Theory},
  vol.~4, no.~4, pp. 29--37, Sep. 1954.

\bibitem{smith2012}
B.~P. Smith, A.~Farhood, A.~Hunt, F.~R. Kschischang, and J.~Lodge, ``Staircase
  {{Codes}}: {{FEC}} for 100 {{Gb}}/s {{OTN}},'' \emph{J. Light. Technol.},
  vol.~30, no.~1, pp. 110--117, Jan. 2012.

\bibitem{shehadeh2025}
M.~Shehadeh, F.~R. Kschischang, A.~Y. Sukmadji, and W.~Kingsford,
  ``Higher-{{Order Staircase Codes}},'' \emph{IEEE Trans. Inf. Theory}, pp.
  1--1, 2025.

\bibitem{pyndiah1998}
R.~Pyndiah, ``Near-optimum decoding of product codes: Block turbo codes,''
  \emph{IEEE Trans. Commun.}, vol.~46, no.~8, pp. 1003--1010, Aug. 1998.

\bibitem{hager2018}
C.~H{\"a}ger and H.~D. Pfister, ``Approaching {{Miscorrection-Free
  Performance}} of {{Product Codes With Anchor Decoding}},'' \emph{IEEE Trans.
  Commun.}, vol.~66, no.~7, pp. 2797--2808, Jul. 2018.

\bibitem{lei2019}
Y.~Lei, B.~Chen, G.~Liga, X.~Deng, Z.~Cao, J.~Li, K.~Xu, and A.~Alvarado,
  ``Improved {{Decoding}} of {{Staircase Codes}}: {{The Soft-Aided
  Bit-Marking}} ({{SABM}}) {{Algorithm}},'' \emph{IEEE Trans. Commun.},
  vol.~67, no.~12, pp. 8220--8232, Dec. 2019.

\bibitem{sheikh2019a}
A.~Sheikh, A.~{Graell i Amat}, and G.~Liva, ``Binary {{Message Passing
  Decoding}} of {{Product-Like Codes}},'' \emph{IEEE Trans. Commun.}, vol.~67,
  no.~12, pp. 8167--8178, Dec. 2019.

\bibitem{miao2022}
S.~Miao, L.~Rapp, and L.~Schmalen, ``Improved {{Soft-Aided Decoding}} of
  {{Product Codes With Dynamic Reliability Scores}},'' \emph{J. Light.
  Technol.}, vol.~40, no.~22, pp. 7279--7288, Nov. 2022.

\bibitem{strasshofer2023}
A.~Stra{\ss}hofer, D.~Lentner, G.~Liva, and A.~{Graell i Amat},
  ``{Soft-Information Post-Processing for Chase-Pyndiah Decoding Based on
  Generalized Mutual Information},'' in \emph{2023 12th International Symposium
  on Topics in Coding (ISTC)}.\hskip 1em plus 0.5em minus 0.4em\relax IEEE,
  2023, pp. 1--5.

\bibitem{deng2023}
S.~Deng, Z.~Xiao, J.~Sha, and Z.~Wang, ``An {{Adaptive Chase-Pyndiah
  Algorithm}} for {{Turbo Product Codes}},'' \emph{IEEE Commun. Lett.},
  vol.~27, no.~4, pp. 1065--1069, Apr. 2023.

\bibitem{yuan2025a}
P.~Yuan, M.~M{\'e}dard, K.~Galligan, and K.~R. Duffy, ``Soft-output ({{SO}})
  {{GRAND}} and {{Iterative Decoding}} to {{Outperform LDPC Codes}},''
  \emph{IEEE Trans. Wireless Commun.}, pp. 1--1, 2025.

\bibitem{chase1972}
D.~Chase, ``Class of algorithms for decoding block codes with channel
  measurement information,'' \emph{IEEE Trans. Inf. Theory}, vol.~18, no.~1,
  pp. 170--182, Jan. 1972.

\bibitem{goalic2002}
A.~Goalic, K.~{Cavalec-Amis}, and V.~Kerbaol, ``Real-time turbo decoding of
  block turbo codes using the {{Hartmann-Nazarov}} algorithm on the {{DSP Texas
  TMS320C6201}},'' in \emph{2002 {{IEEE Int}}. {{Conf}}. {{Commun}}.}, vol.~3,
  Apr. 2002, pp. 1716--1720 vol.3.

\bibitem{ashikhmin2004}
A.~Ashikhmin and S.~Litsyn, ``Simple {{MAP}} decoding of first-order
  {{Reed-Muller}} and {{Hamming}} codes,'' \emph{IEEE Trans. Inf. Theory},
  vol.~50, no.~8, pp. 1812--1818, Aug. 2004.

\bibitem{nazarov1998}
L.~Nazarov and V.~Smolyaninov, ``Use of fast {{Walsh-Hadamard}} transformation
  for optimal symbol-by-symbol binary block-code decoding,'' \emph{Electron.
  Lett.}, vol.~34, no.~3, pp. 261--262, Feb. 1998.

\end{thebibliography}

\begin{table*}[htp]
    \centering
    \caption{Optimized parameters for each half iteration for decoding of the $(256,247,4)^2$ extended Hamming \ac{PC}.}
    \begin{NiceTabular}{r|c|C{.95cm}C{.95cm}ccccc}
    \toprule
    & Decoder 
    & \multicolumn{2}{c}{Optimized CP \cite{strasshofer2023}} 
    &  \multicolumn{2}{c}{$\ac{SOCS}(\beta)$} 
    & $\ac{SOCS}(\Ball_1(\Test))$ 
    & $\ac{SOCS}(\Ball_1(\List))$ 
    & $\ac{SOCS}(\Test)$
    \\
    & Design $E_\mathrm{b}/N_0$ 
    & \multicolumn{2}{c}{$4.55\,\mathrm{dB}$} 
    & \multicolumn{2}{c}{$4.40\,\mathrm{dB}$} 
    & $4.30\,\mathrm{dB}$ 
    & $4.30\,\mathrm{dB}$ 
    & $4.30\,\mathrm{dB}$
    \\
    \cmidrule(lr){3-4} 
    \cmidrule(lr){5-6} 
    \cmidrule(lr){7-7} 
    \cmidrule(lr){8-8} 
    \cmidrule(lr){9-9}
    & Parameters & $\alpha$ & $\beta$ & $\alpha$ & $\beta$ & $\alpha$ & $\alpha$ & $\alpha$\\
    \midrule
    \multirow{7}{*}{\STAB{\rotatebox[origin=c]{90}{Half iteration}}} & 1 & 0.44 & 0.35 & 0.41 & $5.0\times 10^{-4}$ & 0.92 & 1.25 & 1.12 \\
    & 2 & 0.56 & 1.55 & 0.49 & $2.5\times 10^{-4}$ & 0.76 & 1.18 & 1.08 \\
    & 3 & 0.55 & 1.95 & 0.46 & $2.5\times 10^{-4}$ & 0.78 & 1.25 & 1.10 \\
    & 4 & 0.62 & 2.85 & 0.52 & $1.3\times 10^{-4}$ & 0.70 & 1.20 & 1.16 \\
    & 5 & 0.64 & 3.75 & 0.54 & $8.9\times 10^{-5}$ & 0.74 & 1.51 & 1.06 \\
    & 6 & 0.75 & 4.95 & 0.66 & $4.5\times 10^{-5}$ & 0.78 & 1.36 & 1.14 \\
    & 7 & 0.94 & 6.25 & 0.67 & $7.9\times 10^{-6}$ & 0.68 &  1.53 & 1.18 \\
    \bottomrule 
\end{NiceTabular}
    \label{tab:parameters_ham}
\end{table*}

\begin{table*}[htp]
    \centering
    \caption{Optimized parameters for each half iteration for decoding of the $(256,239,6)^2$ extended \ac{BCH} \ac{PC}.}
    \begin{NiceTabular}{r|c|C{.95cm}C{.95cm}ccccc}
    \toprule
    & Decoder 
    & \multicolumn{2}{c}{Optimized CP \cite{strasshofer2023}} 
    &  \multicolumn{2}{c}{$\ac{SOCS}(\beta)$} 
    & $\ac{SOCS}(\Ball_2(\Test))$ 
    & $\ac{SOCS}(\Ball_2(\List))$ 
    & $\ac{SOCS}(\Test)$
    \\
    & Design $E_\mathrm{b}/N_0$ 
    & \multicolumn{2}{c}{$3.80\,\mathrm{dB}$} 
    & \multicolumn{2}{c}{$3.80\,\mathrm{dB}$} 
    & $3.65\,\mathrm{dB}$ 
    & $3.65\,\mathrm{dB}$ 
    & $3.65\,\mathrm{dB}$
    \\
    \cmidrule(lr){3-4} 
    \cmidrule(lr){5-6} 
    \cmidrule(lr){7-7} 
    \cmidrule(lr){8-8} 
    \cmidrule(lr){9-9}
    & Parameters & $\alpha$ & $\beta$ & $\alpha$ & $\beta$ & $\alpha$ & $\alpha$ & $\alpha$\\
    \midrule
    \multirow{7}{*}{\STAB{\rotatebox[origin=c]{90}{Half iteration}}} 
    & 1 & 0.17 & 0.21 & 0.22 & $3.2\times 10^{-7}$ & 0.88 & 1.10 & 1.00 \\
    & 2 & 0.24 & 0.53 & 0.28 & $2.1\times 10^{-7}$ & 0.86 & 1.10 & 1.08 \\
    & 3 & 0.28 & 0.73 & 0.32 & $1.4\times 10^{-7}$ & 0.76 & 1.06 & 1.14 \\
    & 4 & 0.31 & 1.29 & 0.38 & $2.6\times 10^{-8}$ & 0.74 & 1.14 & 1.04 \\
    & 5 & 0.38 & 1.33 & 0.42 & $1.7\times 10^{-8}$ & 0.86 & 1.34 & 1.44 \\
    & 6 & 0.47 & 1.89 & 0.58 & $1.4\times 10^{-9}$ & 0.82 & 1.48 & 1.26 \\
    & 7 & 0.59 & 2.81 & 0.64 & $7.9\times 10^{-10}$ & 0.84 & 1.44 & 1.30 \\
    \bottomrule 
\end{NiceTabular}
    \label{tab:parameters_bch}
\end{table*}

\end{document}